\documentclass[aps,prd,onecolumn,groupedaddress,showpacs,nofootinbib,amssymb
]{revtex4}
\usepackage[dvips]{graphicx}
\usepackage{amssymb}
\usepackage{amsmath}
\usepackage{graphicx}
\usepackage{amsfonts}
\usepackage{bm}

\begin{document}

\title{Constraints on Singular Evolution from Gravitational Baryogenesis}
\author{V.~K.~Oikonomou,$^{1,2}$\,\thanks{v.k.oikonomou1979@gmail.com}}
\affiliation{ $^{1)}$ Tomsk State Pedagogical University, 634061 Tomsk, Russia\\
$^{2)}$ Laboratory for Theoretical Cosmology, Tomsk State University of Control Systems
and Radioelectronics (TUSUR), 634050 Tomsk, Russia\\
}

\begin{abstract}
We investigate how the gravitational baryogenesis mechanism can potentially constrain the form of a Type IV singularity. Specifically, we study two different models with interesting phenomenology, that realize two distinct Type IV singularities, one occurring at the end of inflation and one during the radiation domination era or during the matter domination era. As we demonstrate, the Type IV singularities occurring at the matter domination era or during the radiation domination era, are constrained by the gravitational baryogenesis, in such a way so that these do not render the baryon to entropy ratio singular. Both the cosmological models we study cannot be realized in the context of ordinary Einstein-Hilbert gravity, and hence our work can only be realized in the context of $F(R)$ gravity and more generally in the context of modified gravity only. 
\end{abstract}


\maketitle



\def\pp{{\, \mid \hskip -1.5mm =}}
\def\cL{\mathcal{L}}
\def\be{\begin{equation}}
\def\ee{\end{equation}}
\def\bea{\begin{eqnarray}}
\def\eea{\end{eqnarray}}
\def\tr{\mathrm{tr}\, }
\def\nn{\nonumber \\}
\def\e{\mathrm{e}}

\section{Introduction}

Singularities frequently occur in physical theories, and these indicate probably the inability of the physical theory to describe the physics successfully. An example is the Coulomb potential which diverges at $r=0$, an issue which is resolved in the context of quantum electrodynamics. In general relativity singularities also occur, and these are classically treated in most cases. These occur in the process of gravitational collapse and are protected by the formation of event horizons. The theoretical description of singularities in general relativity was firstly done by Hawking and Penrose \cite{hawkingpenrose}, and in most cases, singularities are related to black holes and these are spacelike singularities. However, in cosmology also timelike singularities occur, in which case these are called finite time singularities, firstly classified in \cite{Nojiri:2005sx}. In the case of finite time singularities, some, or all, physical quantities that can be defined on a constant $t$ spacelike three dimensional hypersurface, diverge. However, there are four different types of finite time singularities which determine the way that the physical quantities diverge. The most severe type of finite time singularities is the Big Rip, and the most ``mild'' from a phenomenological point of view is the Type IV singularity, see \cite{Nojiri:2005sx}. The less destructive types of singularities were firstly studied in \cite{barrownew}, where the types of singularities later called sudden, were studied. As it was shown in \cite{barrownew} closed universes obeying $\rho + 3p>0$ and $\rho>0$ need not to recollapse, because a pressure singularity could occur, before these Universes achieved an expansion maximum. This was a counter-example to what was believed at the time and it was the first discovery of these types of singularity in cosmology with an explicit construction given in Ref. \cite{barrownew}. Sudden singularities were further studied in \cite{Barrow:2004xh1,Barrow:2004xh2} and \cite{barrow1,barrow2,barrow3,barrow4,barrow5,barrow6,barrow7,barrow8}. The Type IV singularity has some phenomenologically appealing applications and implications in cosmology, as was shown in \cite{Barrow:2015ora,noo1,noo2,noo212,noo2a,noo2b,noo3,noo4,noo5,noo6,noo7}. Particularly, it possible to incorporate a Type IV singularity in an inflationary $F(R)$ model, without the physical quantities being divergent at some time instance. However, the implications of the Type IV singularity appear at the dynamical evolution of the cosmological theory. Particularly, as was shown in \cite{noo4,noo5}, the dynamics of the cosmological evolution is abruptly interrupted at the Type IV singularity, which indicates that the theory is dynamically unstable. This was interpreted as an indicator of graceful exit in some $F(R)$ inflationary models \cite{noo4,noo5}. Also, the same occurs for a bounce cosmology, in which case the instability indicates that the bouncing cosmology is unstable at the bouncing point \cite{noo4}. The above hold true in the case that the singularity occurs at the end of inflation, and in most cases the standard inflationary dynamical evolution is abruptly interrupted.

However, the Type IV singularity can occur at any cosmic time instance of cosmological evolution, for example during the matter domination or during the radiation domination era. During inflation, in principle there is no severe constraint that might restrict the Type IV singular evolution, see \cite{noo4,noo5} for an account on singular inflation and also for some review on inflation, see \cite{inflation1,inflation2,inflation3,inflation4,inflation5,inflation6,pert1,pert2,pert3}. The purpose of this article is to investigate how the singular evolution is restricted from gravitational baryogenesis, if the singularities occur during the matter domination or the radiation domination era. We shall present two models which contain Type IV singularities occurring in the matter and radiation era. In addition, both models will be singular at the end of inflation too, where also a Type IV singularity will occur. Both the models are phenomenologically interesting since the late-time acceleration, the early-time acceleration and the radiation domination or the matter domination eras, can simultaneously be described by a single singular cosmological model, which is also compatible with the Planck collaboration observational data \cite{planck1,planck2}. We shall briefly describe the two models, which will be further studied in more detail in a future work. As we shall demonstrate, the process of gravitational baryogenesis can impose some restrictions on the Type IV singularities, if the latter occur during the matter or radiation domination eras. Gravitational baryogenesis refers to a mechanism that materializes the excess of matter over anti-matter, which might occur during the matter or the radiation eras. Particularly, the gravitational baryogenesis mechanism and terminology was firstly presented in Ref. \cite{baryog1} and later further developed in \cite{baryog2,baryog3,baryog4,baryog5,baryog6}. According to the gravitational baryogenesis mechanism, a $\mathcal{C}\mathcal{P}$-violating interaction that generates the observed baryon asymmetry in the Universe, is realized by the presence of the following effective higher order term,
\begin{equation}\label{baryonassterm}
\frac{1}{M_8^2}\int \mathrm{d}^4x\sqrt{-g}(\partial_{\mu} R) J^{\mu}\, ,
\end{equation}
where $M_*$ denotes the cutoff scale of the underlying effective theory, which is responsible for the term (\ref{baryonassterm}) and also $R$, $g$,  and $J^{\mu}$, stand for the Ricci scalar, the trace of the metric, and the baryonic matter current. As we explicitly demonstrate, a Type IV singularity in the cosmological evolution, severely affects the baryogenesis process, due to the existence of the term (\ref{baryonassterm}), via the term $\dot{R}$, if the singularity occurs during the matter or radiation domination era. Hence, our focus is to study all the possible cases that this may occur, and constraint the Type IV singular evolution, so that no inconsistencies occur with regards to the gravitational baryogenesis.

The outline of the paper is as follows: In section I, we briefly review some basic information regarding the gravitational baryogenesis mechanism and also some essential information for the finite time singularities. In section II we present a singular model of cosmological evolution, in the context of which, early-time acceleration, late-time acceleration and the radiation domination era can be described simultaneously by using a single model. We investigate which vacuum $F(R)$ gravity can generate such evolution, emphasizing in the radiation domination era, and we determine how the gravitational baryogenesis constraints this model. In section III we examine another singular model, in which case, early-time acceleration, late-time acceleration and the matter domination era are described by this model, and we also investigate how the gravitational baryogenesis process constraints the singular model. The conclusions follow in the end of the paper.

\section{Gravitational Baryogenesis and Finite Time Singularities Essentials}

\subsection{Gravitational Baryogenesis}

The present time observations indicate an excess of matter over antimatter, with this excess being supported by Big Bang nucleosynthesis successful predictions \cite{bb1}, and the Cosmic Microwave Background observational data \cite{bb2}. In addition, the absence of radiation originating from matter-anti-matter annihilation interactions, also strongly supports the excess of matter over antimatter. The prediction for the baryon to entropy ratio is $\frac{\eta_B}{s}\simeq 9.2\times 10^{-11}$, and it still remains a mystery why this baryon to entropy ratio exists in the Universe. According to the Sakharov criteria \cite{sakharov}, baryon asymmetry in the Universe can occur if one of the following statements holds true,  
\begin{itemize}
    \item There exist baryon number violating particle interactions.
    \item There exist $\mathcal{C}$ and $\mathcal{C}\mathcal{P}$ violating particle interactions.
    \item The thermodynamical processes are non-equilibrium thermodynamical processes.
\end{itemize}    
As it was shown in Ref. \cite{baryog1}, a $\mathcal{C}\mathcal{P}$-violating interaction term has the following form,
\begin{equation}\label{interactioncpvio}
\frac{1}{M_8^2}\int \mathrm{d}^4x\sqrt{-g}(\partial_{\mu} R) J^{\mu}\, ,
\end{equation}
with the parameter $M_*$ denoting the cutoff scale of the higher order underlying effective theory, that generates the term (\ref{interactioncpvio}). Moreover, $J^{\mu}$, $R$ and $g$, stand for the baryonic matter current, the Ricci scalar and the trace of the metric tensor respectively. The interaction term (\ref{interactioncpvio}) frequently occurs in the context of quantum gravity or supergravity theories, see for example \cite{baryog1} and references therein. Also in an expanding Universe, the presence of the interaction term (\ref{interactioncpvio}), generates the violation of the $\mathcal{C}\mathcal{P}\mathcal{T}$ symmetry, and in effect, baryon--anti-baryon asymmetry occurs. A very profound assumption is that in order for the interaction term to generate the baryon asymmetry, the existence of a thermal equilibrium is assumed. Effectively, in the process of the Universe's expansion, after the temperature drops below a critical temperature $T_D$, the asymmetry that remains, is approximately equal to \cite{baryog1},
\begin{equation}\label{baryontoentropy}
\eta_B=\frac{n_B}{s}\simeq \frac{\dot{R}}{M_*^2 T}\Big{|}_{T_D}\, ,
\end{equation}
with $T_D$ denoting the temperature at which the baryon violating processes take place. In addition, note that we denoted with $\eta_B$, the baryon to entropy ratio.

\subsection{Finite-time Singularities and Background Geometry Conventions}

Before providing some basic information about the finite time timelike singularities that occur in cosmology, let us briefly present the geometric framework that will be assumed throughout this paper. The geometric background is assumed to be a flat Friedmann-Robertson-Walker (FRW) one, with line element,
\be
\label{metricfrw} ds^2 = - dt^2 + a(t)^2 \sum_{i=1,2,3}
\left(dx^i\right)^2\, ,
\ee
where $a(t)$ denotes the scale factor. In addition, it is assumed that the connection is an affine connection, which is symmetric, metric compatible and torsion-less, the so-called Levi-Civita connection. Moreover, the Ricci scalar corresponding to the metric (\ref{metricfrw}) has the following form,
\begin{equation}
\label{ricciscal}
R=6\left(2H(t)^2+\dot{H}(t)\right)\, ,
\end{equation}
with $H(t)$ denoting the Hubble rate, $H(t)=\dot{a}/a$. Obviously, if the Ricci scalar has the form (\ref{ricciscal}), it is obvious that the only surviving term of the interaction (\ref{interactioncpvio}), is the one proportional to $\dot{R}$. So effectively, we shall be interested in the calculation of $\dot{R}$. This can also be seen from Eq. (\ref{baryontoentropy}).

Now let s briefly recall some essential information on the finite time singularities. These were firstly classified in a systematic way in Ref.~\cite{Nojiri:2005sx}, and the classification was made by taking into account the following physical quantities, the effective energy density, the effective pressure and the scale factor, and only in one case, the higher derivatives of the Hubble rate $H(t)$. According to Ref. \cite{Nojiri:2005sx} there exist four different types of singularities, the Type I, II, III, IV, which are listed in detail below:
\begin{itemize}
\item Type I (``The Big Rip Singularity''): It is a crushing type singularity, for which all the physical quantities which can be defined on a constant cosmic time $t$, three dimensional spacelike hypersurface, strongly diverge. So, the pressure, the energy density and the scale factor diverge, as the time approaches the singularity instance. or details with regards to this singularity, consult \cite{ref51,ref52,ref53,ref54,ref55,ref56}.
\item Type II (``Sudden Singularity''): This kind of singularities are known as sudden singularities, and were studied in Ref. \cite{Barrow:2004xh1}. In this case, only the pressure diverges, while the scale factor and the energy density remain finite.
\item Type III: This type of singularity is the second most crushing singularity, after the Big Rip, and in this case, both the pressure and energy density diverge, while the scale factor is finite.
\item Type IV: This is the most harmless singularity, due to the implied phenomenology. Actually, as was shown in \cite{noo4,noo5}, the physical quantities are finite, and the effect of the singularity appears in the parameters that determine the dynamical evolution of the cosmological system, like the slow-roll parameters. In this case, all the aforementioned quantities are finite, but only the higher derivatives of the Hubble rate may diverges, that is $\frac{\mathrm{d}^{n}H}{\mathrm{d}t^n}\rightarrow \infty $, with $n\geq 2$. 
\end{itemize}
In this paper we shall be interested on the Type IV singularity and we shall investigate the constraints on the Type IV singular cosmological evolution, imposed by the gravitational baryogenesis term of Eq. (\ref{interactioncpvio}).

\section{Singular Evolution Unifying Early-time, Late-time Acceleration and Radiation Domination Era}
 
As we mentioned in the previous sections, when the singularity occurs during or at the end of the inflationary era, there is no constraint imposed from the gravitational baryogenesis mechanism. However, if the Type IV singularity occurs during the radiation era or during the matter domination era, this can potentially cause inconsistencies in the gravitational baryogenesis mechanism. In this section we shall explicitly demonstrate the inconsistencies that a Type IV singularity generates, if the singularity occurs during the matter domination era. To this end, consider the following model, which unifies the late-time acceleration, radiation domination and early-time acceleration,
\begin{equation}\label{singgenerradiation}
H(t)=\frac{1}{2 \left(\frac{1}{ H_0}+t\right)}+\e^{-(t-t_s)^{\gamma }} 
\left(\frac{H_0}{2}+H_i (t-t_i)\right)+f_0 (t-t_r)^{\delta } (t-t_s)^{\gamma 
}\, ,
\end{equation}
where $H_0$, $t_i$ and $H_i$ being parameters related to the Starobinsky \cite{starobinsky1,starobinsky2} $R^2$ inflation model, see \cite{noo4}. As it can be seen, the model of Eq. (\ref{singgenerradiation}) contains two Type IV singularities occurring at $t=t_s$ and at $t=t_r$ respectively, so both the parameters $\gamma$ and $\delta$ are assumed to satisfy $\delta,\gamma >1$. In addition, the Type IV singularity that occurs at $t=t_s$ is assumed to occur near the end of the inflationary era, while the other singularity occurs much more later, and during the radiation domination era, at $t=t_r$. The model of Eq. (\ref{singgenerradiation}) has quite appealing features, since at early times, near $t\sim t_s$, the Hubble rate becomes approximately, 
\begin{equation}\label{approx}
H(t)\simeq H_0+H_i(t-t_i)\, ,
\end{equation}
since the term $\sim \frac{1}{2 \left(\frac{1}{H_0}+t\right)}$ is approximately equal to $\sim \frac{H_0}{2}$, when $t\simeq t_s$. Also, the term that contains the two Type IV singularities, that is $\sim 
(t-t_s)^{\gamma}(t-t_r)^{\delta }$, becomes zero near $t\sim t_s$. Thereby, the early time behavior of the model resembles the behavior of the Starobinsky $R^2$ gravity, since the Hubble rate (\ref{approx}) is identical to the $R^2$ inflation model \cite{starobinsky1,starobinsky2,noo4}. Hence, the model at early times predicts an accelerating era, which produces a scale invariant spectrum, compatible with present time observations \cite{planck1,planck2}. Correspondingly, when $t\simeq t_r$, the model of Eq. (\ref{singgenerradiation}) describes the radiation domination era, since the Hubble rate becomes approximately equal to,
\begin{equation}\label{reds}
H(t)\simeq \frac{1}{2 \left(\frac{1}{ H_0}+t\right)}\, ,
\end{equation}
which exactly describes the radiation domination era. Accordingly, the late-time behavior of the model (\ref{singgenerradiation}) in terms of the Hubble rate is,
\begin{equation}\label{reds1}
H(t)\simeq f_0 (t-t_r)^{\delta } (t-t_s)^{\gamma }\simeq f_0 (t)^{\delta+\gamma }\, ,
\end{equation}
which describes an acceleration era which occurs at late-times. In order to further support our claims, let us calculate the effective equation of state (EoS) for the model (\ref{singgenerradiation}), and we investigate how this becomes for the various eras we mentioned above. The EoS in the context of modified gravity description for the model (\ref{singgenerradiation}), is equal to,
 \begin{align}
\label{generaleosformodstarmod1}
&w_{\mathrm{eff}}=-1-\frac{2 \left(e^{-(t-t_s)^{\gamma }} H_i-\frac{1}{2 \left(\frac{1}{H_0}+t\right)^2}-e^{-(t-t_s)^{\gamma }} \left(\frac{H_0}{2}+H_i (t-t_i)\right) (t-t_s)^{-1+\gamma } \gamma \right)}{3 \left(\frac{1}{2 \left(\frac{1}{H_0}+t\right)}+e^{-(t-t_s)^{\gamma }} \left(\frac{H_0}{2}+H_i (t-t_i)\right)+f_0 (t-t_r)^{\delta } (t-t_s)^{\gamma }\right)^2} \\ \notag &
+\frac{2\left(f_0 (t-t_r)^{\delta } (t-t_s)^{-1+\gamma } \gamma +f_0 (t-t_r)^{-1+\delta } (t-t_s)^{\gamma } \delta \right)}{3 \left(\frac{1}{2 \left(\frac{1}{H_0}+t\right)}+e^{-(t-t_s)^{\gamma }} \left(\frac{H_0}{2}+H_i (t-t_i)\right)+f_0 (t-t_r)^{\delta } (t-t_s)^{\gamma }\right)^2}
\end{align}
Thereby, the behavior of the EoS for cosmic times near the first Type IV singularity is,
\begin{equation}\label{eosearlyts1}
w_{\mathrm{eff}}\simeq -1-\frac{2 \left(-\frac{H_0^2}{2}+H_i\right)}{3 (H_0+H_i (t-t_i))^2}\, ,
\end{equation}
which describes a nearly de Sitter acceleration era, since $w_{\mathrm{eff}}\simeq -1$. At cosmic times near the radiation domination era, the EoS becomes approximately equal to,
\begin{equation}\label{eosearlyts11}
w_{\mathrm{eff}}\simeq 
-1-\frac{2 \left(-\frac{H_0^2}{2}+H_i\right)}{3 (H_0+H_i (t-t_i))^2}\, ,
\end{equation}
and since $t\sim t_0\gg t_s$, in conjunction with $t\gg \frac{1}{ H_0}$, the expression in Eq. (\ref{eosearlyts11}) is simplified as follows, 
\begin{equation}\label{eosearlyts1a1}
w_{\mathrm{eff}}\simeq -1-\frac{2 \left(-\frac{1}{2 (t)^2}\right)}{3 \left(\frac{1}{2 (t)}\right)^2}\, .
\end{equation}
where we took into account the fact that for $t\gg 1$, for $t$ belonging to the radiation era, the terms containing the exponential are suppressed exponentially, and therefore, the EoS becomes, 
\begin{equation}\label{eosearlyts21}
w_{\mathrm{eff}}\simeq \frac{1}{3}\, ,
\end{equation}
which clearly describes the radiation domination era. Finally, at late times, the EoS becomes approximately equal to,
\begin{equation}\label{eosearlyts31}
w_{\mathrm{eff}}\simeq -1-\frac{2 \left(f_0 (t-t_r)^{\delta } (t-t_s)^{-1+\gamma } \gamma +f_0 (t-t_r)^{-1+\delta } (t-t_s)^{\gamma } \delta \right)}{3 \left(f_0 (t-t_r)^{\delta } (t-t_s)^{\gamma }\right)^2}\, ,
\end{equation}
which can be further simplified as follows,
\begin{equation}\label{eosearlyts41}
w_{\mathrm{eff}}\simeq -1-\frac{2 t^{-1-\gamma -\delta } (\gamma +\delta )}{3 f_0}\, ,
\end{equation}
where we took into account that $t\gg 1$, in conjunction with the fact that  $\gamma,\delta>1$. Hence, the late-time behavior is described by a nearly de Sitter acceleration. After describing in brief the behavior of the model (\ref{singgenerradiation}), let us present which vacuum $F(R)$ gravity can generate the evolution near the radiation domination era, that is near $t\simeq t_r$. For reviews on the vast subject of $F(R)$ gravity we refer to \cite{reviews1,reviews1a,reviews11,reviews12,reviews13}. By using the reconstruction techniques of Refs. \cite{Nojiri:2006gh,Capozziello:2006dj,sergbam08}, we easily find $F(R)$ gravity that can generate the cosmological evolution of Eq. (\ref{reds}), which is,
\begin{equation}\label{det}
F(R)\simeq -\frac{3 C_2 H_0^2 \left(18 H_0^2+R+\sqrt{R} \sqrt{12 H_0^2+R}\right)}{\left(\sqrt{R}-\sqrt{12 H_0^2+R}\right)^2}\, .
\end{equation}
The details of this calculation shall be presented in a future work related to the cosmological behavior of the model (\ref{singgenerradiation}) for all cosmic times.

Having described the model with emphasis near the Type IV singularity occurring during the radiation domination era, let us investigate how the gravitational baryogenesis constraints the second Type IV singularity. Recall that the singularity type of the Hubble rate (\ref{singgenerradiation}) is given below,
\begin{itemize}\label{lista}
\item $\gamma,\delta<-1$ corresponds to the Type I singularity.
\item $-1<\gamma,\delta<0$ corresponds to Type III singularity.
\item $0<\gamma,\delta<1$ corresponds to Type II singularity.
\item $\gamma,\delta>1$ corresponds to Type IV singularity.
\end{itemize}
 So in order to have two Type IV singularities, it is required that $\gamma, \delta >1$. However, this might cause inconsistencies in the baryon to entropy ratio $\eta_B$ of Eq. (\ref{baryontoentropy}). Recall that the baryon to entropy ratio is $\eta_B\sim \dot{R}$, so it is determined by the first derivative of the Ricci scalar (\ref{ricciscal}). A direct calculation of the Ricci scalar (\ref{ricciscal}) corresponding to the Hubble rate (\ref{singgenerradiation}) yields,
\begin{align}\label{dotr}
& R(t)=6 e^{-(t-t_s)^{\gamma }} H_i-\frac{6}{2 \left(\frac{1}{H_0}+t\right)^2}+12 \left(\frac{1}{2 \left(\frac{1}{H_0}+t\right)}+e^{-(t-t_s)^{\gamma }} \left(\frac{H_0}{2}+H_i (t-t_i)\right)+f_0 (t-t_r)^{\delta } (t-t_s)^{\gamma }\right)^2 \\ \notag & -6 e^{-(t-t_s)^{\gamma }} \left(\frac{H_0}{2}+H_i (t-t_i)\right) (t-t_s)^{-1+\gamma } \gamma +f_0 (t-t_r)^{\delta } (t-t_s)^{-1+\gamma } \gamma +6 f_0\, . (t-t_r)^{-1+\delta } (t-t_s)^{\gamma } \delta\, .
\end{align}
Correspondingly, a direct calculation of $\dot{R}$ yields,
\begin{align}\label{dotyielddirecctcal}
& \dot{R}(t)=\frac{6}{\left(\frac{1}{H_0}+t\right)^3}-12 e^{-(t-t_s)^{\gamma }} H_i (t-t_s)^{-1+\gamma } \gamma \\ \notag & -6e^{-(t-t_s)^{\gamma }} \left(\frac{H_0}{2}+H_i (t-t_i)\right) (t-t_s)^{-2+\gamma } (-1+\gamma ) \gamma +f_0 (t-t_r)^{\delta } (t-t_s)^{-2+\gamma } (-1+\gamma ) \gamma \\ \notag &
+6 e^{-(t-t_s)^{\gamma }} \left(\frac{H_0}{2}+H_i (t-t_i)\right) (t-t_s)^{-2+2 \gamma } \gamma ^2+12 f_0 (t-t_r)^{-1+\delta } (t-t_s)^{-1+\gamma } \gamma  \delta +6f_0 (t-t_r)^{-2+\delta } (t-t_s)^{\gamma } (-1+\delta ) \delta\\ \notag & +24 \left(\frac{1}{2 \left(\frac{1}{H_0}+t\right)}+e^{-(t-t_s)^{\gamma }} \left(\frac{H_0}{2}+H_i (t-t_i)\right)+f_0 (t-t_r)^{\delta } (t-t_s)^{\gamma }\right) \times \\ \notag &
\Big{(}e^{-(t-t_s)^{\gamma }} H_i-\frac{1}{2 \left(\frac{1}{H_0}+t\right)^2}-e^{-(t-t_s)^{\gamma }} \left(\frac{H_0}{2}+H_i (t-t_i)\right) (t-t_s)^{-1+\gamma } \gamma +f_0 (t-t_r)^{\delta } (t-t_s)^{-1+\gamma } \gamma \\ \notag & +f_0 (t-t_r)^{-1+\delta } (t-t_s)^{\gamma } \delta \Big{)}\, .
\end{align}
From Eq. (\ref{dotyielddirecctcal}), it is obvious that when $1<\gamma <2$, the function $\dot{R}$ and therefore the baryon to entropy ratio, strongly diverge at $t=t_r$, since terms $\sim (t-t_r)^{\delta-2}$ appear in (\ref{dotyielddirecctcal}). Therefore, in order that no inconsistencies occur in the gravitational particle production, we must require that $\delta >2$. This is the main result of this paper and the result implies that if the gravitational baryogenesis mechanism is realized in nature, the Type IV singularity may have catastrophic consequences for the baryon to entropy ratio, if $1<\gamma <2$. Hence, since the baryon to entropy ratio is a small quantity and it is well defined in the present Universe, this implies that either the Type IV singularity during the radiation era does not exist, or that it exists and the parameter $\delta$ is chosen to satisfy $\delta>2$. As we demonstrate in the next section, the same constraint applies if the Type IV singularity occurs during the matter domination era. To this end, we shall use a similar model to the one appearing in Eq. (\ref{singgenerradiation}).

\section{Singular Evolution Unifying Early-time, Late-time Acceleration and Matter Domination Era}

The same constraints we found in the previous section also apply in the case that the singularity occurs during the matter domination era. A viable model that unifies the matter domination era, with the late-time and early-time acceleration, is described by the following Hubble rate,
\begin{equation}\label{singgener}
H(t)=\frac{2}{3 \left(\frac{4}{3 H_0}+t\right)}+\e^{-(t-t_s)^{\gamma }} 
\left(\frac{H_0}{2}+H_i (t-t_i)\right)+f_0 (t-t_0)^{\delta } (t-t_s)^{\gamma 
}\, ,
\end{equation}
with the parameters $t_s$, $H_0$, $t_0$, $\gamma$, $\delta $, $H_i$, $f_0$ and $t_i$, being chosen as in the previous section. Also $t_0$ is assumed to be a time instance during the matter domination era. If $\gamma, \delta>1$, then the cosmological evolution described by (\ref{singgener}) develops two Type IV singularities, one in the end of the inflationary era, at $t=t_s$ and one during the matter domination era, at $t=t_0$. The model has the same characteristics as the model we described in the previous section, appearing in Eq. (\ref{singgenerradiation}). So at early times, it behaves as the Starobinsky $R^2$ inflation model, with the EoS being approximately equal to,
\begin{equation}\label{eosearlyts}
w_{\mathrm{eff}}\simeq -1-\frac{2 \left(\frac{3 H_0}{4}+H_i\right)}{3 (H_0+H_i 
(t-t_i))^2}\, ,
\end{equation}
which describes nearly de Sitter acceleration, since $H_0$ and $H_i$ are chosen to satisfy $H_0, H_i\gg 1$. During the matter domination era, the EoS becomes,
\begin{equation}\label{eosearlyts2}
w_{\mathrm{eff}}\simeq -1-\frac{2 \left(-\frac{2}{3 t^2}\right)}{3 
\left(\frac{2}{3 t}\right)^2}= 0\, ,
\end{equation}
and therefore our claim that the era is the matter domination one, is perfectly justified since $w_{\mathrm{eff}}\simeq 0$. Moreover, at late times, the EoS behaves,
\begin{equation}\label{eosearlyts4}
w_{\mathrm{eff}}\simeq -1-\frac{2 t^{-1-\gamma -\delta } \gamma }{3 
f_0}-\frac{2 t^{-1-\gamma -\delta } \delta }{3 f_0}\, .
\end{equation}
Hence the late-time behavior is described by a nearly de Sitter acceleration era. Both the models we studied in this paper have quite appealing phenomenology and will be studied in detail elsewhere. Now we are interested on the constraints imposed on the Type IV singularity which occurs during the matter domination era. As in the previous section, the term $\dot{R}$, relevant for gravitational baryogenesis, behaves as follows,
\begin{align}\label{dotyielddirecctcal1}
& \dot{R}(t)=\frac{8}{\left(\frac{4}{3 H_0}+t\right)^3}-12 e^{-(t-t_s)^{\gamma }} H_i (t-t_s)^{-1+\gamma } \gamma +6 f_0 (t-t_0)^{\delta } (t-t_s)^{-2+\gamma } (-1+\gamma ) \gamma\\ \notag &
-6 e^{-(t-t_s)^{\gamma }} \left(\frac{H_0}{2}+H_i (t-t_i)\right) (t-t_s)^{-2+\gamma } (-1+\gamma ) \gamma +6e^{-(t-t_s)^{\gamma }} \left(\frac{H_0}{2}+H_i (t-t_i)\right) (t-t_s)^{-2+2 \gamma } \gamma ^2\\ \notag &
+12 f_0 (t-t_0)^{-1+\delta } (t-t_s)^{-1+\gamma } \gamma  \delta +6f_0 (t-t_0)^{-2+\delta } (t-t_s)^{\gamma } (-1+\delta ) \delta\\ \notag &
+24 \left(\frac{2}{3 \left(\frac{4}{3 H_0}+t\right)}+e^{-(t-t_s)^{\gamma }} \left(\frac{H_0}{2}+H_i (t-t_i)\right)+f_0 (t-t_0)^{\delta } (t-t_s)^{\gamma }\right) \times \\ \notag &
\Big{(} e^{-(t-t_s)^{\gamma }} H_i-\frac{2}{3 \left(\frac{4}{3 H_0}+t\right)^2}+f_0 (t-t_0)^{\delta } (t-t_s)^{-1+\gamma } \gamma\\ \notag &
-e^{-(t-t_s)^{\gamma }} \left(\frac{H_0}{2}+H_i (t-t_i)\right) (t-t_s)^{-1+\gamma } \gamma +f_0 (t-t_0)^{-1+\delta } (t-t_s)^{\gamma } \delta \Big{)} \,
\, .
\end{align}
Hence in this case too, in order for the baryon to entropy ratio to be finite, the parameter $\delta$ has to satisfy $\delta>2$. Thereby, we demonstrated that although a Type IV singularity has no effect on physical quantities, if this occurs at the end of the inflationary era, it may severely modify the physics of the matter and radiation era, if the singularity occurs at this era. Particularly, the gravitational baryogenesis mechanism imposes some constraints on the form of the Type IV singularity, which if it's form is realized in the Hubble rate as $\sim (t-t_f)^{\delta }$, then the parameter $\delta$ must satisfy $\delta>2$.

Before we close this section, an important comment is in order. Both the cosmological models (\ref{singgenerradiation}) and (\ref{singgener}) cannot be realized by the ordinary Einstein-Hilbert gravity, and can only be realized in the context of modified gravity, for example $F(R)$ gravity, and example of which we gave in Eq. (\ref{det}). In a future work, we shall give explicit expressions of the $F(R)$ gravities which realize the cosmologies (\ref{singgenerradiation}) and (\ref{singgener}).

\section{Conclusions}

In this paper we investigated a particular case for which the form of the Type IV singularity is constrained by a physical process. Particularly, the gravitational baryogenesis predicts that the baryon to entropy ratio is analogous to $\dot{R}$, and this term can be singular for a Type IV singularity. Since the baryon to entropy ratio is measured and its value is known, this means that the singularity form must definitely be constrained. But we should put things into proper context, and therefore we should note that only when the Type IV singularity occurs during the radiation domination era or during the matter domination era, it is possible that the baryon to entropy ratio might be problematic in the case of a Type IV singularity. Actually, as is already shown in the literature, the appearance of a Type IV singularity does not have a direct effect on the physical quantities that can be defined on a three dimensional spacelike hypersurface defined by the time instance that the singularity occurs. Specifically, the scale factor, the energy density and the pressure are finite for a Type IV singularity, and singularities appear only in the higher derivatives of the Hubble rate. As was shown in \cite{noo4,noo5}, if the Type IV singularity occurs at the end of the inflationary era, this leads to severe instabilities in the slow-roll parameters, which in turn indicate that the system is dynamically unstable. But all the physical quantities are perfectly finite and no singularity occurs in these. However, in this paper we showed a specific example which can impose constraints on the Type IV singularity. Specifically, we studied two models, which have interesting qualitative features, since in the first model, three different eras are described by the same model, the early-time era, the late-time era and also the radiation domination era, while in the second model, the unification of early and late-time acceleration with the matter domination era is achieved. In the first model, two Type IV singularities occur, one at the end of the inflationary era and one during the radiation domination era. The one occurring during the radiation domination era is constrained by the gravitational baryogenesis mechanism, since it depends on the term $\dot{R}$, and the latter depends on $\sim (t-t_r)^{\delta-2}$, which is singular if $1<\delta <2$. The same considerations apply for the second model, only that in this case, the radiation domination era is replaced by the matter domination era. We need to note that both the cosmological scenarios we investigated cannot be realized by ordinary Einstein-Hilbert gravity, so these can be realized only in the context of modified gravity, like for example $F(R)$ gravity. In a future paper we shall present a detailed account and description of these two interesting models, and work is in progress.

\section*{Acknowledgments}

This work is supported by Min. of Education and Science of Russia (V.K.O).

\end{document}